\documentstyle[aps,prl,multicol,graphicx]{revtex}

\begin{document}
\draft


\title{Dissipationless Spin Transport in Thin Film Ferromagnets}

\author{J\"urgen K\"onig,$^{1,2,3}$ 
Martin Chr. B{\o}nsager,$^{3,}$\cite{bonsager} 
and A. H. MacDonald$^{2,3}$}

\address{
$^1$Institut f\"ur Theoretische Festk\"orperphysik, Universit\"at Karlsruhe, 
76128 Karlsruhe, Germany\\
$^2$Department of Physics, University of Texas at Austin, Austin, 
TX 78712\\
$^3$Department of Physics, Indiana University, Bloomington, IN 47405}

\date{\today}

\maketitle

\begin{abstract}

Metallic thin film ferromagnets generically possess spiral states that carry 
dissipationless spin currents. 
We relate the critical values of these supercurrents to micromagnetic material 
parameters, identify the circumstances under which the supercurrents will be 
most robust, and propose experiments which could reveal this new collective 
transport behavior.

\end{abstract}



\begin{multicols}{2}

In ferromagnetic metals and semiconductors quasiparticle states can be 
manipulated by external magnetic fields that couple to the 
spin-magnetization-density collective coordinate. 
This property is responsible for related robust magnetoresistance effects that 
occur in various geometries such as anisotropic magnetoresistance in bulk 
samples, giant magnetoresistance \cite{gmrrefs1,gmrrefs2} in metallic 
multilayers, and tunnel magnetoresistance \cite{tmrrefs1,tmrrefs2} in tunnel 
junctions. 
In this paper we propose a distinctly different type of spin-dependent
transport effect, 
in which spin current is carried collectively rather than by quasiparticles. 
Because the spin current is non-zero when its quasiparticles are in 
equilibrium, it is carried without dissipation.  
This spin-supercurrent state occurs only in easy-plane ferromagnets and will 
be robust only when anisotropy within the easy plane is weak.
We propose an experiment to observe this effect in thin films of 
ferromagnetic metals.

The key observation that motivates this proposal arises by considering the 
class of excited states obtained from the ferromagnetic ground state by 
following its adiabatic evolution as equal and opposite constant vector 
potentials are introduced for up and down spins, with the spin-quantization 
axis perpendicular to the ferromagnet's easy plane. 
The many-particle Hamiltonian is
\begin{equation}
   {\cal H} = \sum_i \left[{\hbar^2 \over 2m} \left( {\bf k}_i + 
        {{\bf Q} \sigma_z \over 2}\right)^2 + v({\bf r}_i) \right]
        + {\cal H}_{\rm el-el},
\label{eq:ham}
\end{equation}
with ${\cal H}_{\rm el-el} = \sum_{i<j} {e^2 / |{\bf r}_i-{\bf r}_j|}$.
The vector potentials for spin $\sigma$ are  
${\bf A}_\sigma = {\bf Q}_\sigma(\hbar c /e)$ with 
${\bf Q}_{\uparrow} = - {\bf Q}_{\downarrow} = {\bf Q} /2$.
They can be removed by gauge transformations, multiplying the many-particle 
wavefunction by $\exp (i {\bf Q}_\sigma \cdot {\bf r})$ for each electron with 
spin $\sigma$.
In a paramagnet, the ground-state wavefunction would, therefore, evolve 
trivially with $\bf Q$ and the ground-state energy would be independent of 
$\bf Q$.
In a ferromagnet, however, a change in the phase relationship between up 
spins and down spins alters the magnetic order and will change the energy.

We start with a ground state that has a spontaneous spin density 
along the $\hat {\bf x}$ direction.
Its magnitude is
$ m({\bf r}) = \langle \Psi_{\uparrow}^{\dagger}({\bf r}) 
\Psi_{\downarrow}({\bf r})\rangle_0$,
where $\Psi_{\sigma}({\bf r})$ is an electron field operator for an electron 
with spin $\sigma=\uparrow, \downarrow$, 
and $\langle ...\rangle_0$ denotes a ground-state expectation value. 
For small $\bf Q$, 
$\Psi_{\sigma}({\bf r}) \to \exp(i {\bf Q}_\sigma \cdot {\bf r})
\Psi_{\sigma}({\bf r})$ and the order parameter rotates in the 
$\hat {\bf x} - \hat {\bf y}$ plane as a function of ${\bf r}$,
\begin{equation}
   \langle{\bf s}({\bf r})\rangle_{\bf Q} = m_{Q} [\cos({\bf Q} \cdot 
        {\bf r})
        \hat {\bf x} + \sin({\bf Q} \cdot {\bf r}) \hat {\bf y}] \, ,
\end{equation}
i.e., it forms a spiral spin state.
As $Q$ increases, the order parameter's spatial dependence will \cite{bydefn} 
cause a decrease in the magnetic condensation energy and in the magnitude of 
the order parameter. 
{\em Dependence of the ground-state energy $E$ on $Q$ implies that these 
many-particle eigenstates have finite current densities with equal magnitude 
and opposite direction for up and down spins},
\begin{equation}
   {\bf j}_{\uparrow} = 
        {c \over V} {\partial E({\bf A}_\uparrow,{\bf A}_\downarrow) \over
        \partial {\bf A}_{\uparrow}} 
        \bigg|_{{\bf A}_\uparrow =-{\bf A}_\downarrow} =
        \frac{e}{\hbar} \frac{\partial \epsilon ({\bf Q})}{\partial {\bf Q}} =
        - {\bf j}_{\downarrow} \, ,
\label{eq:curdens}
\end{equation}
where $\epsilon ({\bf Q})$ is the energy per unit volume.

As we discuss at greater length below, easy-plane anisotropy ascribes 
a topological character to the wavevector $\bf Q$, so that these currents 
can decay only by phase slip processes that have large barriers, 
{\it i.e., these are dissipationless supercurrents}.

Equation~(\ref{eq:curdens}) is similar to the connection between the 
exchange-coupling of ferromagnets separated by a tunnel junction and the 
spin currents that flow between them \cite{slonczewski}.
Our proposal for supercurrents in ferromagnets is related to Anderson's 
discussion of superconductivity \cite{anderson} in terms of magnetic order in 
an effective spin model; the physics of the two ordered states appears similar 
if a particle-hole transformation is made in one of the spin subspaces.  
The supercurrents we propose are also related to those supported in 
double-layer quantum Hall systems
\cite{qheold1,qheold2,qheold3,jordan,jordan2,abolfath},
where ordered states form that are describeable either as pseudospin 
ferromagnets or as electron-hole pair condensates \cite{eh_pair1,eh_pair2}.   
In fact, the role of magnetic anisotropy in 
controlling the observability of these supercurrents is connected in 
part with the role of band hybridization terms in controlling the 
observability of collective electron-hole-pair transport 
in excitonic insulators \cite{exc1,exc2,exc3,exc4}.

To illustrate these ideas we consider the simplest possible
microscopic model of a metallic ferromagnet, a fermion gas with
delta-function repulsive particle-particle interactions 
$U \delta ({\bf r}_i - {\bf r}_j)$ treated in a mean-field approximation. 
The unrestricted Hartree-Fock Hamiltonian for the ordered state with 
wavevector ${\bf Q} = Q\hat {\bf x}$ is
\end{multicols}
\begin{eqnarray}
   {\cal H}^{HF} = {Vh^2\over U} +
        \sum_{\bf k}
        \left( c_{{\bf k} + {\bf Q}/2,\uparrow}^\dagger \, \, 
                c_{{\bf k} - {\bf Q}/2,\downarrow}^\dagger
        \right)
        \left(
        \begin{array}{cc}
                \epsilon_{{\bf k} + {\bf Q}/2} & -h \\
                -h & \epsilon_{{\bf k} - {\bf Q}/2}
        \end{array} \right)
        \left(
        \begin{array}{c}
                c_{{\bf k} + {\bf Q}/2,\uparrow}\\
                c_{{\bf k} - {\bf Q}/2,\downarrow}
        \end{array} \right)
   = {Vh^2\over U} + \sum_{{\bf k},\pm} E_{{\bf k},\pm}
        a^{\dagger}_{{\bf k},\pm} a_{{\bf k},\pm} 
\label{eq:mft}
\end{eqnarray}
\begin{multicols}{2}
\noindent
where $\epsilon_{\bf k} = \hbar^2k^2/(2m)$, the ordered-state quasiparticle
energies are
$ E_{{\bf k},\pm} = [ \epsilon_{{\bf k} + {\bf Q}/2} + 
\epsilon_{{\bf k} - {\bf Q}/2} \pm \sqrt{(\epsilon_{{\bf k} + {\bf Q}/2} -
\epsilon_{{\bf k} - {\bf Q}/2})^2 + 4 h^2}]/2$,
%
and the effective magnetic field which splits the quasiparticle bands by 
$\pm h$ with $h=Um_Q$ is determined self-consistently by
\begin{equation}
  {h\over U} = {1\over V}{\sum_{\bf k}} ^{\prime}
        \frac{h}{\sqrt{(\epsilon_{{\bf k} + {\bf Q}/2} 
        - \epsilon_{{\bf k} - {\bf Q}/2})^2 + 4 h^2}}.
\label{eq:selfconsistent}
\end{equation}
The prime on the sum in Eq.~(\ref{eq:selfconsistent}) indicates restriction 
to those wavevectors for which only the lower-energy quasiparticle state 
is occupied.

\begin{figure}
\narrowtext
\centerline{\includegraphics[width=8cm]{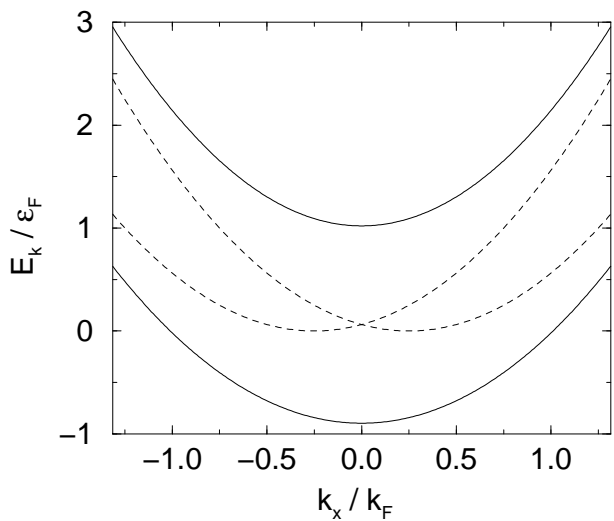}}
\caption{Quasiparticle bands $E_{k,+}$ (upper solid curve) and $E_{k,-}$ 
(lower solid curve) for $Q=0.5 k_F$, $UD(\epsilon_F)=1.5$ and $k_y=k_z=0$.
For comparison we also show the dispersion $\epsilon_{{\bf k} + {\bf Q}/2}$ and
$\epsilon_{{\bf k} - {\bf Q}/2}$ for zero order parameter (dashed lines).}
\label{fig:one}
\end{figure}

\begin{figure}
\vspace*{1.4cm}
\centerline{\includegraphics[width=8cm]{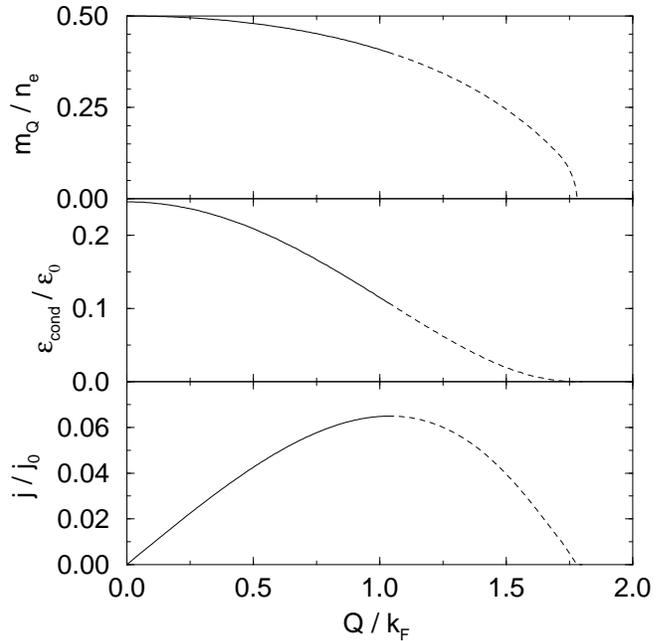}}
\vspace*{.4cm}
\caption{The order parameter $m_Q$ normalized to electron density $n_e$, the 
magnetic condensation-energy density $\epsilon_{\rm cond}$ normalized to the 
energy density of the disordered state, $\epsilon_0=(3/5) n_e \epsilon_F$, 
and the spin supercurrent density $j=j_\uparrow = -j_\downarrow$ normalized 
to $j_0 = en_e\hbar k_F/m$ as a function of the ordering wavevector $Q$ for 
$UD(\epsilon_F)=1.5$.
The dashed lines indicate an instability regime against phase separation into 
regions with larger and smaller $Q$.}
\label{fig:two}
\end{figure}

The procedure described above provides a mean-field approximation to 
the ferromagnet spin-supercurrent states. 
In Fig.~\ref{fig:one} we plot quasiparticle bands for a typical
model of this type at a moderately large value of $Q=0.5 k_F$, where 
$k_F$ is the Fermi wavevector for zero order parameter, and $\epsilon_F$ 
is the Fermi energy \cite{comment}.
The product $UD(\epsilon_F)$ was chosen to be close to experimental values 
for Co and Fe (taken from Ref.~\cite{skomskicoey}).
In this case the Stoner criterion \cite{stoner} for mean-field 
ferromagnetism, $UD(\epsilon_F) > 1$, is satisfied.
In Fig.~\ref{fig:two} we plot the order parameter $m_Q$, the magnetic
condensation energy $\epsilon_{\rm cond}$, and the spin supercurrent density 
$j$ as a function of the ordering wavevector $Q$.
Note that the current density is proportional to the derivative of the 
condensation energy in agreement with the more general discussion above.

Our calculations demonstrate that spin supercurrents are possible 
in states with equilibrium quasiparticle populations; 
elastic scattering from occupied to unoccupied states 
cannot provide the current decay mechanism familiar from the 
standard theory of metallic transport.  
To establish the stability of the spin currents it is, however, still 
necessary to show that the spin-supercurrent state is stable against
infinitesimal distortions of its order-parameter field. 
In what follows we demonstrate that magnetic anisotropy is necessary for 
stability.
Since real metallic ferromagnets are much more complex than the toy model 
system discussed above, we now turn to a phenomenological approach that will
allow us to relate critical currents to known micromagnetic parameters.

We consider a generalized Landau-Ginzburg model for the dependence of an 
easy-plane ferromagnet's free-energy density \cite{skomskicoey,landau} on its 
magnetic state:
\begin{equation}
   f = - |\alpha| {\bf M} \cdot {\bf M} + \frac{\beta}{2} ({\bf M} \cdot
        {\bf M})^2 + \tilde A |{\bf \nabla M}|^2 + \tilde K M_z^2.
\label{eq:gl}
\end{equation}
The free energy of this model is minimized by a constant magnetization 
in the $\hat {\bf x}-\hat {\bf y}$ easy-plane with magnitude 
$M_0= \sqrt{|\alpha|/\beta}$.
The resulting dependence of energy density on magnetization orientation at 
fixed magnitude allows us to identify $A = \tilde A M_0^2$ with the 
exchange constant and $K = \tilde K M_0^2$ with the uniaxial anisotropy 
coefficient of the ferromagnet's micromagnetic energy 
functional \cite{skomskicoey}. 
The spin-supercurrent state has magnetization 
${\bf M}_Q(x) = M_Q (\cos(Qx),\sin(Qx),0)$, where 
$M_Q^2=(|\alpha|-\tilde A Q^2)/\beta$ is decreasing with $Q$ as in our 
microscopic calculations.
Using Eq.~(\ref{eq:curdens}), we find that the spin supercurrent density is
\begin{equation}
   j(Q) = \frac{2 e \tilde A Q}{\hbar \beta} (|\alpha| - \tilde A Q^2),
\label{eq:glcurrent}
\end{equation}
reaching a maximum at $Q_{ph}$ where $\tilde A Q_{ph}^2 = |\alpha|/3$.

Expanding around the spin-supercurrent state free energy extremum,
we find that
\begin{eqnarray}
   \delta f &=& 2 \beta M_Q^2 M_{a}^2 +
        \tilde A |{\bf \nabla} M_{a}|^2 + \tilde A |{\bf \nabla} M_{ph}|^2
\nonumber \\
        &&+ 2 Q \tilde A (M_{a} \partial_x M_{ph} - M_{ph} \partial_x M_{a} )
\nonumber \\ 
        &&+ (\tilde K - \tilde A Q^2) M_{z}^2 +
        \tilde A |{\bf \nabla} M_{z}|^2 ,
\label{eq:flucenergy}
\end{eqnarray}
where $M_{a}$ and $M_{ph}$ are the amplitude and phase fluctuations
of the easy-plane magnetization (the projections along and perpendicular 
to ${\bf M}_Q$), while $M_{z}$ is the hard-axis fluctuation. 
The translationally invariant kernel of this quadratic form has three
wavevector $({\bf p})$ dependent eigenvalues:
\begin{eqnarray}
   K_{\pm} &=& \beta M_Q^2 + \tilde A p^2 \pm \sqrt{ \beta^2
        M_Q^4 + 4 {\tilde A}^2 Q^2 p_x^2} 
\label{eq:stability xy}
\\ 
        K_{zz} &=& \tilde K - \tilde A Q^2 + \tilde A p^2.
\label{eq:stability z}
\end{eqnarray}
It follows from Eq.~(\ref{eq:stability xy}) that the spin-supercurrent state 
is stable against easy-plane fluctuations provided that $Q$ is smaller 
than $Q_{ph}$; at larger values of $Q$, energy can be lowered by phase 
separation into regions with larger and smaller $Q$. 
For the soft ferromagnets we have in mind, however, it is the out-of-plane 
fluctuations, described by Eq.~(\ref{eq:stability z}), that become unstable
first. 
For $Q > Q_{z} = \sqrt{K/A}$, the spin supercurrent can relax by 
tilting out of the easy-plane to one of the poles and unwinding phase with no 
energy cost.
In Table~\ref{table:01} we list $Q_z$ values and the corresponding critical 
current densities $j_{\rm crit} = j(Q_z)$ for some common soft thin film 
magnets, including only the shape (magnetostatic) contribution 
$K_{\rm shape}=\mu_0 M_0^2/2$ to $K$.
From our model calculation and the results shown in Fig.~\ref{fig:two}
we conclude that $Q_{ph}$ is typically of the order of the Fermi wavevector
$k_F$, i.e., much larger than $Q_z$.
To estimate $j_{\rm crit}$ we can, therefore, linearize 
Eq.~(\ref{eq:glcurrent}) in $Q$, 
\begin{equation}
   j_{\rm crit} = 2{e\over\hbar} A Q_z = 2{e\over\hbar} 
        \sqrt{AK_{\rm shape}} \, ,
\end{equation}
to obtain the large critical currents listed 
in Table~\ref{table:01}.

We have so far neglected magnetocrystalline anisotropy, since it is much 
weaker than shape anisotropy in the situation we have in mind.  
It does, however, break rotational symmetry within the easy plane and has the 
tendency to fix the phase and, thus, to suppress the supercurrents.
When an in-plane anisotropy term is included in the energy-density functional, 
extrema at small phase winding rates consist of weakly coupled solitons in
which the magnetization goes from one in-plane minima to another.
($Q = \theta N_s/L$ where $\theta$ is the angle between in-plane minima and 
$N_s/L$ is the soliton density.)  
The energy density at small $Q$ is proportional to the number of solitons.
As a consequence, the minimum spin-current density $j_{\rm min}$, that can be 
supported by a spin-supercurrent state is non-zero.  
To estimate $j_{\rm min}$ for cubic materials we include the leading-order 
bulk cubic anisotropy \cite{skomskicoey} in the energy density,
$K_1^{(c)} \sin^2 \varphi \cos^2 \varphi$
where $\varphi$ is the angle of the order parameter within the easy plane.
For small $Q$ the functional is minimized by a kink soliton solution.
By evaluating the energy of in-plane solitons of this model, we find from 
Eq.~(\ref{eq:curdens}) that 
\begin{equation}
   {j_{\rm min}\over j_{\rm crit}} = {1\over 4\pi} 
        \sqrt{K_1^{(c)}\over K_{\rm shape}} 
\end{equation}
which is of the order of $1.5 \%$ (see Table~\ref{table:01}).
$\langle 100 \rangle$ hcp Cobalt thin films with in-plane easy-axis will 
typically have still smaller values of $j_{\rm min}$ because of the higher 
hexagonal symmetry.  
From these considerations, we conclude that spin supercurrents will be 
observable at moderate current densities only in materials that have weak  
magnetic anisotropy within the easy plane.
Because of their extremely weak magnetocrystalline anisotropies, homogeneous
permalloy samples might be ideal candidates for the experiments proposed below.
Although the course grained in-plane magnetic ansiotropy can in principle
be fine-tuned to zero, spin-rotational invariance in the easy-plane will
always be broken by disorder terms in the microscopic Hamiltonian.
Since dissipationless spin supercurrents will not occur if these disorder terms 
are too strong, the effects we propose are more likely to be observable in 
homogeneous alloys.

\begin{figure}
\centerline{\includegraphics[width=8.5cm]{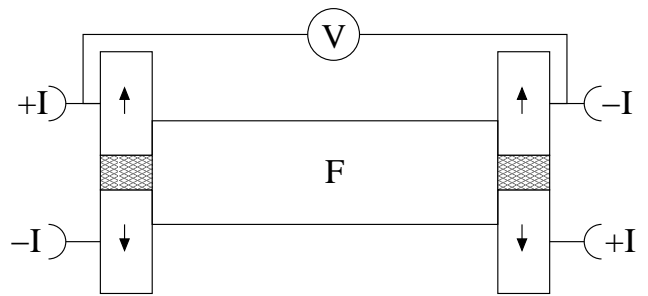}}
\caption{Schematic illustration of one possible experimental set up to
prepare a spin-supercurrent state.}
\label{fig:three}
\end{figure}

One possible experimental arrangement in which this collective transport 
phenomena could be detected is illustrated schematically in 
Fig.~\ref{fig:three}.
An easy-plane thin film ferromagnet (F) is connected to four spin-selective
leads (full spin polarization in the leads is optimal but not required)
that feed opposing up and down spin currents, where ``up'' and ``down'' refers 
to the direction {\it perpendicular} to the thin film.
We emphasize that even with recent advances in transition metal ferromagnet 
spintronics, realizing a system with this geometry represents an experimental 
challenge.
In this setup, a quasiparticle current would flow dissipatively between upper 
and lower leads on both the left and right hand side of the thin film 
ferromagnet.
A sizeable voltage drop (measured, e.g., between the upper leads), proportional 
to the injected currents, would result.
Its exact value depends on the resistivity of the ferromagnet and on details of 
the geometry and is not a concern here.
If the collective transport effect predicted here occurred, however, currents
with opposite spin would flow without a voltage drop across the sample, from
left to right and vice versa.
Dissipationless current flow in the bulk could still be masked by resistance
in the film-lead contacts or by collective spiral wave phase-slip processes.
Our uncertainty in the magnitude of the contact resistances compared to the
quasiparticle resistance makes our proposal somewhat speculative.
A collective element to the spin transport could be unambiguously
identified by driving the critical current density $j$ through either the
maximum or the minimum current, $j_{\rm crit}$ or $j_{\rm min}$, or by
reversing the spin orientations of the leads on one side of the sample.
The later change would have no effect on the measured voltage if the current
were carried entirely by quasiparticles but would increase the voltage if part
of the current was carried collectively.

In conclusion, we have examined circumstances under which dissipationless spin
supercurrents, associated with spiral magnetic order, can occur in thin film 
ferromagnets.
We have estimated critical values of these supercurrents and proposed an
experiment to generate and detect this new collective transport behavior.

We acknowledge helpful discussions with Jack Bass, James Erskine, and Leonard 
Kleinmann, and support by the Deutsche Forschungsgemeinschaft, by the Welch 
Foundation, and by the Indiana 21st Century Fund.

\begin{table}
\begin{tabular}{l|ccc}
   & Fe & Co & Ni\\
\hline
$\mu_0M_0 \, [{\rm T}]$ & 2.15 & 1.81 & 0.62\\
$A \, [{\rm pJ \, m^{-1}}]$ & 8.3 & 10.3 & 3.4\\
$K_1^{(c)} \, [{\rm MJ} \, {\rm m}^{-3}]$ & 0.048 & & $-0.005$\\
$Q_z \, [{\rm nm^{-1}}]$ & 0.47 & 0.36 & 0.21\\
$j_{\rm crit} \, [{\rm A\, cm}^{-2}]$ & $1.19 \times 10^{9}$ & 
$1.11 \times 10^{9}$ & $2.2 \times 10^{8}$ \\
$j_{\rm min}/j_{\rm crit}$ & 0.013 & & 0.015
\end{tabular}
\caption{Saturation moment $\mu_0M_0$, exchange constant $A$, cubic  
anisotropy constant $K_1^{(c)}$, critical wavevector $Q_z$, critical spin 
current density $j_{\rm crit}$, and ratio of minimum to critical spin current 
density for common soft thin film magnets.
The values for $\mu_0M_0$, $A$, and $K_1^{(c)}$ are taken from 
Ref.~\protect\cite{skomskicoey}.}
\label{table:01}
\end{table}
\end{multicols}


\begin{thebibliography}{99}

\bibitem[*]{bonsager} 
Present address: Seagate Technology, 7801 Computer Avenue South, Bloomington, 
MN 55435

\bibitem{gmrrefs1} 
M.N. Baibich {\it et al.}, Phys.\ Rev.\ Lett. {\bf 61}, 2472 (1988).

\bibitem{gmrrefs2} 
G. Binasch, P. Gr\"unberg, F. Saurenbach, and W. Zinn, Phys.\ Rev.\ B {\bf 39}, 
4828 (1989).

\bibitem{tmrrefs1} 
M. Julliere, Phys.\ Lett. {\bf 54A}, 225 (1975).

\bibitem{tmrrefs2} 
J.S. Moodera, L.R. Kinder, T.M. Wong, and R. Meservey, Phys.\ Rev.\ Lett.
{\bf 74}, 3273 (1995).

\bibitem{bydefn} 
If the ground state is a unidirectional spin-density
wave the minimum of $\epsilon(Q)$ will occur at a finite value of
$Q$, rather than at $Q=0$.

\bibitem{slonczewski}
J.C. Slonczewski, Phys. Rev. B {\bf 39}, 6995 (1989).

\bibitem{anderson}
P.W. Anderson, Phys.\ Rev. {\bf 112}, 1900 (1958).

\bibitem{qheold1} 
X.-G. Wen and A. Zee, Phys.\ Rev.\ Lett. {\bf 69}, 1811 (1992).

\bibitem{qheold2} 
K. Yang {\it et~al.}, Phys.\ Rev.\ Lett. {\bf 72}, 732 (1994).

\bibitem{qheold3} 
K. Moon {\it et~al.}, Phys.\ Rev.\ B {\bf 51}, 5138 (1995).

\bibitem{jordan} 
J. Kyriakidis, D. Loss, and A.H. MacDonald, Phys.\ Rev.\ Lett. {\bf 83}, 
1411 (1999); {\bf 85}, 2222 (2000).

\bibitem{jordan2} 
J. Kyriakidis and L. Radzihovsky, cond-mat/0010329.

\bibitem{eh_pair1}
Yu.E. Lozovik and V.I. Yudson, Zh.\ Eksp.\ Teor.\ Fiz. {\bf 71}, 738 (1976) 
[Sov.\ Phys.\ JETP {\bf 44}, 389 (1976)].

\bibitem{eh_pair2}
S.I. Shevchenko, Sov.\ J.\ Low.\ Temp.\ Phys. {\bf 2}, 251 (1976).

\bibitem{exc1}
L.V. Keldysh and Yu.V. Kopaev, Sov.\ Phys.\ Solid State {\bf 6}, 2219 (1965).

\bibitem{exc2}
D. J{\'e}rome, T.M. Rice, and W. Kohn, Phys.\ Rev. {\bf 158}, 462 (1967).

\bibitem{exc3}
B.I. Halperin and T.M. Rice, Rev.\ Mod.\ Phys. {\bf 40}, 755 (1968).

\bibitem{exc4}
R.R. Guseinov and L.V. Keldysh, Zh.\ Eksp.\ Teor.\ Fiz. {\bf 63}, 2255 (1972) 
[Sov.\ Phys.\ JETP {\bf 36}, 1193 (1973)].

\bibitem{comment}
We find that there is no excitation gap in the quasiparticle spectrum, similar
to, e.g., $d$-wave superconducters, which show gapless superconductivity.

\bibitem{skomskicoey}
R. Skomski and J.M.D. Coey, {\it Permanent Magnetism} (Institute of Physics
Publishing, Bristol, 1999).

\bibitem{stoner}
E.C. Stoner, Proc.\ Roy.\ Soc. {\bf A154}, 656 (1936).

\bibitem{landau}
L.D. Landau and E.M. Lifshitz, {\it Course of Theoretical Physics}, Vol. 9.

\end{thebibliography}
\end{document}